# Quark Fragmentation into $^3P_J$ Quarkonium

J.P. Ma

Recearch Center for High Energy Physics
School of Physics
University of Melbourne
Parkville, Victoria 3052
Australia

**Abstract**:
We calculate the functions of parton fragmentation into $^3P_J$ quarkonium at order $\alpha_s^2$, where the parton can be a heavy or light quark. The obtained functions explicitly satisfy the Altarelli-Parisi equation and they are divergent, behaving as $z^{-1}$ near $z=0$. However, if one choses the renormalization scale as twice of the heavy quark mass, the fragmentation functions are regular over the whole range of $z$.

# 1. Introduction

   Fragmentation functions are in general nonperturbative objects in the context of the factorization theorem in QCD[1]. This makes it hard to study them by starting directly from QCD. However, if partons, i.e., quarks and glouns, undergo fragmentation into a quarkonium, fragmentation functions can be factorized—they are sums of products of constants and coefficient functions. The constants represent the nonperturbative effects and may be defined as matrix elements in nonrelativestic QCD(NRQCD) or are related to a wavefunction, while the coefficients functions can be calculated with perturbative theory. With this factorization, fragmenation functions for various quarkonia were studied in [2–9].

   A quarkonium contains a heavy quark $Q$ and its antiquark $\bar{Q}$, which move with a small velocity $v$ in the quarkonium rest frame. Because of the small velocity the bound state effect, i.e., the nonperturbative effect in the quarkonium, can be well described by employing NRQCD. Recently, such an approach has been established[10]. The approach is basically distinct than early treatment within the color-singlet model(for a nice review see [11]). In the color-singlet model one treats a quarkonium system simply as a bound state of $Q$ and $\bar{Q}$, where the $Q\bar{Q}$ pair is in a color-singlet state, and the nonperturbative effect is contained in the wavefunction of the bound state. Then an expansion in the small parameter $v$ is made and to leading order only the wavefunction at the origin or its derivative at the origin is involved. This model has serious problems. First, it does not tell us how to handle the Coulomb singularities. Since an expansion in $v$ is made, this type of singularity must appear because of massless photons and massless gluons. The effect related to the Coulomb singularities is nonperturbative. In the color-singlet model these singularities were absorbed into the wavefunction without solide reasons from theories. Second, infrared(I.R.) singularities appear when a $P$-wave quarkonium is involved and appear even at the leading order of $\alpha_s$. In the model the I.R. singularities were regularized as divergences in the limit of the zero-binding energy. Such I.R. singularities clearly indicate that the wavefunction of a $P$-wave quarkonium can not contain all the nonperturbative effects. In the approach[10], quarkonium systems are analysed in the framework of QCD. A systematice expansion in $v$ can be made and the nonperturbative effect is represented by matrix elements in NRQCD. With these matrix elements one can show that Coloumb singularities are factorized into these matrix elements. From the point of view of a relativistice quantum theory a quarkonium system consists not only of a $Q\bar{Q}$ color-singlet but also many other components like $|Q\bar{Q}G> \cdots$ etc.. In the approach of [10] the effect of these other components is also included in the systmatic expansion in $v$. It should be emphasized [10,12] that a $Q\bar{Q}$ color-octet state is as important as a $Q\bar{Q}$ color-singlet for a $P$-wave quarkonium. Therefore one should take both states into account. It is already shown that results at the one-loop level for gluon fragmenation into $^3P_J$ quarkonium[9] are free from I.R. singularities and from Coloumb singularities.

In this work we will study the quark fragmentation into a $^3P_J$ $Q\bar{Q}$ quarkonium at leading order $\alpha_s$. We will show that because of the contribution of the color-octet $Q\bar{Q}$ pair, a light quark $q$ can also fragment into the quarkonium at the same order of $\alpha_s$ as a heavy quark $Q$. The heavy quark fragmentation into $P$-wave quarkonium was originally studied in[8]. As pointed out in [9] the results for charmonium can not be correct, because the fragmenation functions obtained there do not satisfy the Altarelli-Parisi equation.

The work is organized as follows: in Sect. 2 we introduce the definition[13] of renormalized quark fragmentation functions and the factorzation[10] forms for quark fragmentation into $^3P_J$ quarkonium. Further details may be found in [10,13]. In Sect.3 we start from the definiton to calculate the heavy quark fragmentation functions. In Sect.4 we calculate the light quark fragmentation function. Finally we discuss and summarise the results of our work in Sect.5. Throughout this work we always assume that the polarization of the quarkonium is not observed. We will use dimensional regularization and work at leading order in the expansion of $v$.

## 2. The Definition of Quark Fragmentation Function and the Factorization Form for $^3P_J$ quarkonium

As we will use dimensional regularization we give the definiton of the quark fragmentation function in $d$ dimensions. To give the definitions for a fragmentaion function it is convenient to work in the light-cone coordinate system. In this coordinate system a $d$-vector $p$ is expressed as $p^\mu = (p^+, p^-, \mathbf{p_T})$, with $p^+ = (p^0 + p^{d-1})/\sqrt{2}$, $p^- = (p^0 - p^{d-1})/\sqrt{2}$. Introducing a vector $n$ with $n^\mu = (0, 1, 0, \cdots, 0) = (0, 1, \mathbf{0_T})$, the fragmentation function for a spinless hadron $H$ or for a hadron without observing its polarization is defined as[13]:

$$D^{(0)}_{H/q}(z) = \frac{z^{d-3}}{4\pi} \int dx^- e^{-iP^+ x^-/z} \frac{1}{3}\text{Tr}_{\text{color}} \frac{1}{2}\text{Tr}_{\text{Dirac}} \{n \cdot \gamma <0|q(0)$$
$$\bar{P}\exp\{-ig_s \int_0^\infty d\lambda n \cdot G^T(\lambda n^\mu)\} a^\dagger_H(P^+, \mathbf{0_T}) a_H(P^+, \mathbf{0_T}) \quad (2.1)$$
$$P\exp\{ig_s \int_{x^-}^\infty d\lambda n \cdot G^T(\lambda n^\mu)\} \bar{q}(0, x^-, \mathbf{0_T})|0>,$$

where $G_\mu(x) = G^a_\mu(x) T^a$, $G^a_\mu(x)$ is the gluon field and $T^a (a=1,\cdots,8)$ are the $SU(3)$-color matrices. The subscript $T$ denotes the transpose. $a^\dagger_H(\mathbf{P})$ is the creation operator for the hadron $H$. For hadrons with nonzero spin the summation over the spin is understood. The definition is a unrenormalized version. Ultraviolet divergences will appear in $D_{H/q}^{(0)}(z)$ and call for renormaization. Following [13] the renormalized gluon fragmenation function can be defined as:

$$D_{H/Q}(z) = D_{H/Q}^{(0)}(z) + \sum_{a=G,q} \int_z^1 \frac{dy}{y} L_a(\frac{y}{z}) D_{H/a}^{(0)}(y). \tag{2.2}$$

Here the summation over all possible partons is understood. The function $L_a(z)$ is chosen so as to cancel the U.V. divergences. In MS scheme $L_a(z)$ takes the form:

$$L_a(z) = \sum_N \frac{1}{\epsilon^N} L_a^{(N)}(z), \tag{2.3}$$

where $\epsilon = 4 - d$. From Eq.(2.2) one can derive the Altarelli-Parisi type evolution equation for the fragmentation function. We will use the modified MS scheme, where $L_a(z)$ is chosen to cancel the terms with $N_\epsilon = \frac{2}{\epsilon} - \gamma + \ln(4\pi)$. The function $D_{H/q}(z)$ is interpreted as the probability of a quark $q$ with momentum $k$ to decay into the hadron $H$ with momentum component $P^+ = zk^+$, it is gauge invariant by definiton. Further, it is also invariant under a Lorentz boost along the moving direction of the hadron and under a rotation with the direction as the rotate axis.

If the hadron is a $^3P_J$ quarkonium, a factorized form for the fragmentation function can be taken. We will use the notation $\chi_J$ for the $^3P_J$ quarkonium. At the leading order of $v$ $D_{H/q}(z)$ can be written according to [10] as:

$$D_{\chi_J/q}(z) = \frac{\hat{D}_1(z,J)}{M^5} <0|O_1^{\chi_J}(^3P_J)|0> + \frac{\hat{D}_8(z)}{M^3} <0|O_8^{\chi_J}(^3S_1)|0>. \tag{2.4}$$

where $\hat{D}_1(z,J)$ and $\hat{D}_8(z)$ are dimensionless and $\hat{D}_8(z)$ is same for all $J$. The operators $O_1^{\chi_J}(^3P_J)$ and $O_8^{\chi_J}(^3P_1)$ are given by:

$$\begin{aligned}
O_8^H(^3S_1) &= \chi^\dagger \sigma_i T^a \psi (a_H^\dagger a_H) \psi^\dagger \sigma_i T^a \chi \\
O_1^H(^3P_0) &= \frac{1}{3} \chi^\dagger (-\frac{i}{2} \overleftrightarrow{\mathbf{D}} \cdot \sigma) \psi (a_H^\dagger a_H) \psi^\dagger (-\frac{i}{2} \overleftrightarrow{\mathbf{D}} \cdot \sigma) \chi \\
O_1^H(^3P_1) &= \frac{1}{2} \chi^\dagger (-\frac{i}{2} \overleftrightarrow{\mathbf{D}} \times \sigma)_i \psi (a_H^\dagger a_H) \psi^\dagger (-\frac{i}{2} \overleftrightarrow{\mathbf{D}} \times \sigma)_i \chi \\
O_1^H(^3P_2) &= \chi^\dagger (-\frac{i}{2} \overleftrightarrow{\mathbf{D}}_{\{i} \sigma_{j\}}) \psi (a_H^\dagger a_H) \psi^\dagger (-\frac{i}{2} \overleftrightarrow{\mathbf{D}}_{\{i} \sigma_{j\}}) \chi,
\end{aligned} \tag{2.5}$$

where $\mathbf{D}$ is the space part of the covariant derivative $D^\mu$ and $\sigma_i (i=1,2,3)$ is the Pauli matrix. The notation $\{ij\}$ means only the symmetric and traceless part of a tensor taken.

In Eq. (2.5) $\psi$ and $\chi^\dagger$ are fields with two components for the heavy quark $Q$ and its antiquark $\bar{Q}$ in NRQCD. $M$ is the mass of the heavy quark. $a_H^+$ is the creation operator for the hadron in its rest frame. The matrix elements in Eq.(2.4) are defined in NRQCD. In Eq.(2.4) the part with $\hat{D}_8$ is the contribution from a color-octet $Q\bar{Q}$ pair in a $^3S_1$ state and the part with $\hat{D}_1(z,J)$ is the contribution from a color-singlet $Q\bar{Q}$ pair in a $^3P_J$ state. We will call them as the color-octet and color-singlet components respectively. The matrix elements represent the nonperturbative effect, while $\hat{D}_1$ and $\hat{D}_8$ can be calculated perturbatively and they should be free from I.R. singularities.

A good method to calculate $\hat{D}_1$ and $\hat{D}_8$ is to use wavefunctions to project out different states from a general $Q\bar{Q}$ pair. At the leading order of $v$ the projection can easily be worked out, details may be found in [14]. We will use a radial wavefunction $R_1(r)$ to project the $^3P_J$ color-singlet $Q\bar{Q}$ state and an octet radial wavefunction $R_8^{(a)}(r)$ to project the $^3S_1$ color-octet $Q\bar{Q}$ state. Calculating with these wavefunctions the l.h.s of Eq.(2.4) and the matrix elements in the r.h.s of Eq.(2.4) we can extract the functions $\hat{D}_1(z,J)$ and $\hat{D}_8(z)$. The results for $\hat{D}_1(z,J)$ and $\hat{D}_8(z)$ are independent on these wavefunctions. At the order of $\alpha_s$ we consider, only the tree-level results for the matrix elements are needed, they are:

$$\begin{aligned} <0|O_1^{\chi_J}(^3P_J)|0> &= \frac{9(2J+1)}{2\pi}|R_1'(0)|^2, \\ <0|O_8^{\chi_J}(^3S_1)|0> &= \frac{3}{8\pi}\sum_c |R_8^{(c)}(0)|^2, \end{aligned} \quad (2.6)$$

where $R_1'(0)$ is the first derivative of $R_1(r)$ at the origin.

### 3. The Heavy Quark Fragmentation Function

From the defintion in Eq.(2.1) we can always decompose the fragmentation function by sandwiching the operator $\sum_X |X><X|$ between $a_H^\dagger$ and $a_H$ as:

$$D_{H/Q}^{(0)}(z) \sim \sum_X \text{Tr}\{n\cdot\gamma T_H^\dagger T_H\}, \quad (3.1)$$

where $T_H$ may be called the fragmentation amplitude for $Q \to H + X$. Here we have the conservation of total momentum only in the +-direction.

### 3.1 The Color-Singlet Component

The color-singlet component receives nonzero contributions at order $\alpha_s^2$. The Feynman diagrams for $T_H$ are given in Fig.1. Because the $Q\bar{Q}$ pair is in a color-singlet state, there are two gluon lines attached to the quark line. Here, there are no divergences, so renormalization is not required. The calculation is complicated. Because of the summation over intermediate states we encountered integrals of the type:

$$\int (\frac{dq_T}{2\pi})^{d-2} (q_T^2 + \frac{(2-z)^2 M^2}{z^2})^{-n}, \text{ for } n = 2,3,4,5. \quad (3.2)$$

Here $\mathbf{q_T}$ is the transversel momentum of the quark as the intermediate state in Fig.1. The integrals are finite and after performing the integrations we can extract:

$$\begin{aligned}
\hat{D}_1(z, J=0) &= \frac{16}{729} \alpha_s^2(\mu) \frac{z(1-z)^2}{(2-z)^8} \\
&\quad \cdot (192 + 384z + 528z^2 - 1376z^3 + 1060z^4 - -376z^5 + 59z^6), \\
\hat{D}_1(z, J=1) &= \frac{64}{729} \alpha_s^2(\mu) \frac{z(1-z)^2}{(2-z)^8} \\
&\quad \cdot (96 - 288z + 496z^2 - 408z^3 + 202z^4 - 54z^5 + 7z^6), \\
\hat{D}_1(z, J=2) &= \frac{128}{3645} \alpha_s^2(\mu) \frac{z(1-z)^2}{(2-z)^8} \\
&\quad \cdot (48 - 192z + 480z^2 - 668z^3 + 541z^4 - 184z^5 + 23z^6).
\end{aligned} \quad (3.3)$$

These results agree with those in [8]. It is interesting to note that there is a common factor $z(1-z)^2(2-z)^{-8}$ for all $J$, whereas there is a common factor $z(1-z)^2(2-z)^{-6}$ for heavy quark fragmentation into $S$-wave quarkonium[6]. Note that the same diagrams in Fig.1 contribute to heavy quark fragmentation into $S$-wave quarkonium. The difference between these two factors are because for $P$-wave quarkonium the derivative of the fragmentation amplitude with the relative momentum between $Q$ and $\bar{Q}$ is involved whereas for $S$-wave quarkonium only the amplitude itself is involved. The appearance of these common factors can be roughly understood by counting the denominators due to the quark- and gluon propagators in the amplitudes and factors from phase-space. A successful model for heavy quark fragmentation was obtained through this way[15]. However, we will see in the next subsection that such a counting rule will be violated due to renormalization.

### 3.2 The Color-Octet Component

For the color-octet componenet, not only the diagrams in Fig.1, but also the diagrams in Fig.2 will contribute The two diagrams in Fig.2 were missing in [8] and they lead to divergences. Instead of integrals in Eq.(3.2), we have:

$$\int (\frac{dq_T}{2\pi})^{d-2} (q_T^2 + \frac{(2-z)M}{z})^{-n}, \text{ for } n = 1,2,3,4. \quad (3.4)$$

The integral with $n = 1$ is ultraviolet divergent, requiring renormalization. For the renormalization we note that the function of gluon fragmentation into $\chi_J$ quarkonium is nonzero at order $\alpha_s$[9,6,7]:

$$D_{\chi_J/G}(z) = \frac{\pi}{24}\alpha_s \delta(1-z) \cdot \frac{1}{M^3} <0|O_8^{\chi_J}(^3S_1)|0> + O(\alpha_s^2). \tag{3.5}$$

Substituting this into Eq.(2.2) for $H = \chi_J$ we can easily chose the function $L_G(y)$ to cancel the divergence. Finally, we obtain the renormalized function $\hat{D}_8(z)$:

$$\begin{aligned}\hat{D}_8(z) = \frac{1}{36}\alpha_s^2(\mu)\Big\{&\frac{1}{z}(1+(1-z)^2)(\ln\big(\frac{\mu^2}{4M^2}\big) - 2\ln(1-\frac{z}{2})) - z \\ &+ \frac{2(1-z)}{9(2-z)^6}(192 - 1184z + 2016z^2 - 1360z^3 + 352z^4 - 14z^5 - 5z^6)\Big\}.\end{aligned} \tag{3.6}$$

Here there is is no common factor like $z(1-z)^2(1-z)^{-6}$. With the results in Eq.(3.6) and in Eq.(3.3) we complete the heavy quark fragmentation function at order $\alpha_s^2$. This function should in general satisfy its evolution equation:

$$\mu \frac{\partial D_{H/Q}(z,\mu)}{\partial \mu} = \sum_q^{N_f} \int_z^1 \frac{dy}{y} P_{Q\to q}(z/y,\mu) D_{H/q}(y,\mu) + \int_z^1 \frac{dy}{y} P_{Q\to G}(z/y,\mu) D_{H/G}(y,\mu). \tag{3.7}$$

The spliting functions $P_{Q\to q}(y,\mu)$ and $P_{Q\to G}(y,\mu)$ are in the one-loop approximation the same as those for parton distributions. Using this fact and the result in Eq.(3.5) we obtain the evolution equation for the quark fragmenation at order of $\alpha_s^2$:

$$\mu \frac{\partial D_{\chi_J/Q}(z,\mu)}{\partial \mu} = \frac{1}{18}\alpha_S^2(\mu)\frac{1}{M^3} <0|O_8^{\chi_J}(^3S_1)|0> \cdot \frac{1+(1-z)^2}{z}. \tag{3.8}$$

Substituting our results into the l.h.s. in Eq.(3.8) one can check that our results are in agreement with this equation. From our result the fragmentation function is divergent as $z^{-1}$ when $z \to 0$. However, this singularity disappears if we chose the renormalization scale $\mu$ as twice of the mass $M$. The same was also found in the gluon fragmentation in [9]. This property is important for possible applications of our result. In practical applications one solves the evolution equation at $\mu$ numerically, where one needs the moments of our result for the fragmentation function at some initial scale $\mu_0$ as the boundary condition. To insure that the perturbative result is a good approximation, one should chose $\mu_0 \sim M$ to avoid large logarithmc terms in higher order. Our result tells that one should chose $\mu_0 = 2M$ to avoid these terms in higher order and to safely calculate the moments in other hand.

## 4. The Light Quark Fragmentation Function

Since a color-octet $Q\bar{Q}$ pair will lead to a contribution to $P$-wave quarkonium production at the leading order of $v$, a light quark $q$ can undergo fragmentation into $\chi_J$ by generating a color-octet $Q\bar{Q}$ through emission of a virtual gluon. Such a process happens at the same order of $\alpha_s$ as the heavy quark fragmentation. The Feynman diagrams for $T_H$ are those in Fig.2, where the quark line attached by the double line is for the light quark $q$. At the leading order of $\alpha_s$ and $v$ the light quark fragmentation function $D_{\chi_J/q}(z)$ can be written:

$$D_{\chi_J/q}(z) = \frac{\hat{D}_{8,q}(z)}{M^3} <0|O_8^{\chi_J}(^3S_1)|0>. \tag{4.1}$$

The color-singlet component only becomes nonzero in higher order than $\alpha_s^2$. The calculation is similar to the previous section. We introduce the notation $y = \frac{m_q}{M}$, where $m_q$ is the mass of $q$. The result for $\hat{D}_{8,q}(z)$ is:

$$\begin{aligned}\hat{D}_{8,q}(z) = \frac{1}{36}\alpha_s^2(\mu)\{&\frac{1}{z}(1+(1-z)^2)\big[\ln\big(\frac{\mu^2}{4M^2}\big) - \ln\big((1-z) + \frac{1}{4}y^2z^2\big)\big] \\ &- z - 2(1-z)(2+y^2)\frac{z}{4(1-z)+y^2z^2}\}.\end{aligned} \tag{4.2}$$

Again the light quark frgamentation function must satify its evolution equation. At order $\alpha_s^2$ this equation is the same in Eq.(3.8). It is easy to check that the function in Eq.(4.1) and (4.2) satisfies the evolution equation. The light quark fragmenation function has the same property near $z = 0$ as the heavy one, i.e., it is divergent as $z^{-1}$ at any remormalizaion scale $\mu$ except when $\mu = 2M$. The light quark mass $m_q$ can be safely neglected. With $m_q = 0$ the function in (4.2) becomes:

$$\hat{D}_{8,q}(z) = \frac{1}{36}\alpha_s^2(\mu)\{\frac{1}{z}(1+(1-z)^2)\big[\ln\big(\frac{\mu^2}{4M^2}\big) - \ln(1-z)\big] - 2z\}. \tag{4.3}$$

For the convenience of later discussions we introduce here some relations between the various matrix elements in Eq.(2.5). In principle these matrix elements have series expansions in $v$ and the leading order is $v^2$. Since we only work at the leading order, the higher order corrections can be neglected. In this case, the matrix elements in Eq.(2.5) are related to each other with a spin factor of $\chi_J$. We introduce two parameters $H_1$ and $H'_8$ as in [10], and the relations can be expressed as:

$$\begin{aligned}<0|O_1^{\chi_J}(^3P_J)|0> &\approx (2J+1)M^4H_1, \\ <0|O_8^{\chi_J}(^3S_1)|0> &\approx (2J+1)M^2H'_8.\end{aligned} \tag{4.4}$$

With these relations the whole set of the quark fragmention functions contains only two unknown parameters, which can only be computed nonperturbatively or extracted from experiments.

## 5. Discussion and Summary

Some useful quantities of parton fragmentation functions are their first moments. These moments allow one to roughly estimate a single hadron production rate though fragmentation, where the rate may be taken as product of a parton production rate and the corresponding first moment, where the summation over different partons is understood. We will give results of the first moments of our fragmentation functions. We denote the first moment as $M(q_f \to \chi_J)$, where $q_f$ stands for $Q$ or $q$. Taking $\mu = 2M$, we obtain:

$$
\begin{aligned}
M(q \to \chi_J) &\approx 0.029(2J+1)\alpha_s^2(2M)\frac{H'_8}{M}, \\
M(Q \to \chi_0) &\approx 0.024\alpha_s^2(2M)\frac{H'_8}{M} + 0.035\alpha_s^2(2M)\frac{H_1}{M}, \\
M(Q \to \chi_1) &\approx 0.072\alpha_s^2(2M)\frac{H'_8}{M} + 0.039\alpha_s^2(2M)\frac{H_1}{M}, \\
M(Q \to \chi_2) &\approx 0.12\alpha_s^2(2M)\frac{H'_8}{M} + 0.015\alpha_s^2(2M)\frac{H_1}{M}.
\end{aligned}
\tag{5.1}
$$

Here we neglect the mass of the light quark. We take the $c$ quark as an example to give some value for the moments. For the value of $H_1$ and $H'_8$ we use the estimates in [12]. Taking $M = m_c = 1.5\text{GeV}$ and $\alpha_s(2m_c) = 0.26$, the value for $M(q \to \chi_{cJ})$ is $3.8 \times 10^{-6}(2J+1)$ and the value for $M(Q \to \chi_{cJ})$ is $2.6 \times 10^{-5}$, $3.4 \times 10^{-5}$ and $2.5 \times 10^{-5}$ for $\chi_{c0}$, $\chi_{c1}$ and $\chi_{c2}$ repectively. The contribution from the color-octet componennt is not negligible. For $\chi_{c2}$ the contribution from the color-octet component is roughly 70% of the heavy quark moment. From these values one can see that the moments for the light quark fragmentation are roughly one order of magnitude smaller than those for the heavy quark. But the contribution from light quarks should not be neglected, especially in a hadron reaction, since the production rate of light quarks as parton may be larger than the production rate of a heavy quark and hence a substantial contribution from light quark fragmentation to the $\chi_J$ production is possible.

With the results here and those in [9] the functions of all possible parton fragmenation into $^3P_J$ quarkonium are calculated at order $\alpha_s^2$. Only two parameters, which represent the nonperturbative effect at the leading order of $v$, are not known precisely. The functions have the general feature that they are divergent as $z^{-1}$ when $z \to 0$. But at $\mu = 2M$ they are regular distributions over the whole range of $z$. The functions also satify the Altarelli-Parisi equation, as expected.

**Acknowledgment:** The author would like to thank Dr. A. Rawlinson for reading the text carefully. This work is supported by Australain Research Council.

**Figure Captions**

Fig.1: The Feynman diagrams for the color-singlet component of the heavy quark fragmentation. The line is for the heavy quark, the wavy line is for gluons. The double line represents the line operator in Eq.(2.1).

Fig.2 The Feynman diagrams for the color-octet component of the heavy and light quark fragmentation.

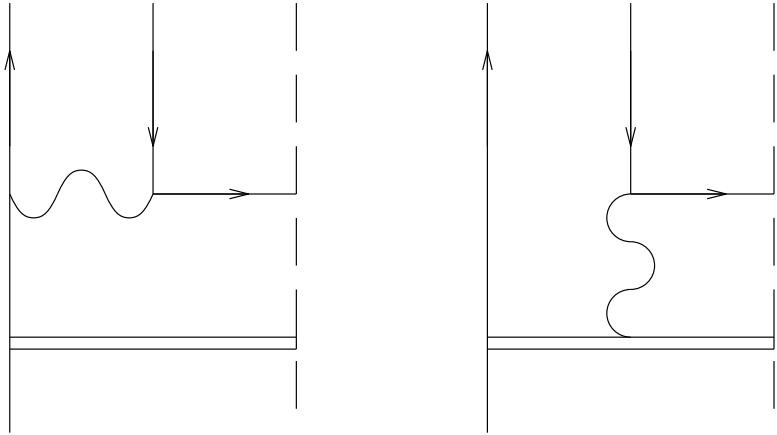

Fig. 1

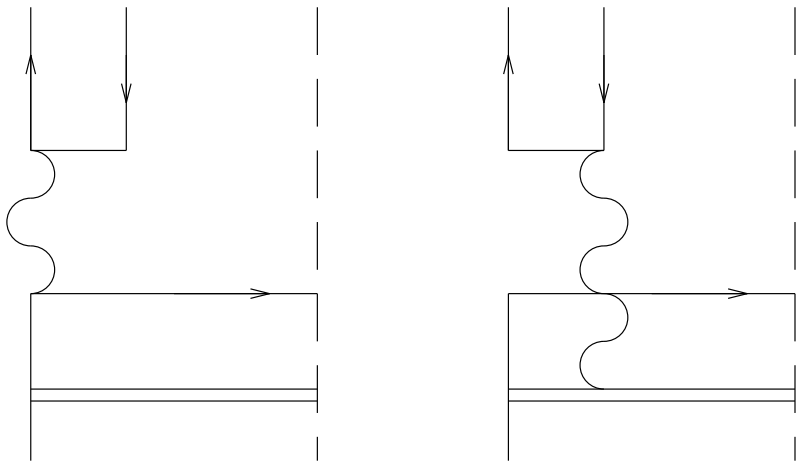

Fig. 2